\documentclass[aps,twocolumn]{revtex4}
\usepackage{graphicx}
\usepackage{dcolumn}
\usepackage{bm}
\usepackage{amssymb}
\usepackage{amsmath}
\usepackage{color}
\usepackage{multirow}
\begin{document}

\title{Hall effect in 2D systems with hopping transport and strong disorder}
\author{A.V. Shumilin$^1$, N.P. Stepina$^2$}
\affiliation{$^1$ Ioffe Institute, 194021 St.-Petersburg, Russia}
\affiliation{$^2$ Institute of Semiconductor Physics, Siberian Branch of the Russian Academy of Sciences, 630090 Novosibirsk, Russia}

\begin{abstract}
We reconsider the theory of Hall effect in the systems with hopping conduction. The purpose of the present study is to compare the percolation approach based on the optimal triad model with numerical simulations and recent experimental results. We show that, in the nearest neighbor hopping regime, the results of the percolation theory agree to the simulation. However, in the variable range hopping (VRH)  regime, the optimal triad model fails to describe the numerical results. It is related to the extremely small probability to find the optimal triad of sites in the percolation cluster in the VRH regime. The contribution of these triads to the Hall effect appears to be small. We describe the Hall mobility in the VRH regime with the empirical law obtained from the numerical results. The law is in agreement with our recent experimental data in 2D quantum dot arrays with the hopping transport.
\end{abstract}

\maketitle

\section{Introduction}

The hopping transport is one of the fundamental kinds of electron transport. It appears in a number of different systems from doped semiconductors to granular metals and organic semiconductors. The Hall effect in metals and semiconductors with free carriers
yields important information on their properties allowing to determine carriers concentration and mobility. However, the understanding of the Hall effect in the hopping regime is far from being complete.  Moreover, there
is still no general agreement on whether the Hall effect
can be observed in the hopping regime at all. The problems exist both in the theoretical and experimental approaches to this phenomenon.

The theoretical study started from the work of T. Holstein \cite{Holstein}. It was shown that although the Hall effect is absent in the model of two-site one-phonon hops that are invoked in most of the hopping transport theories, the Hall effect exists due to many-phonon processes.  The magnetic field-dependent
 contribution to the hopping probability arises from the interference between the amplitude of direct transition between the initial and the final sites of the hop and the  amplitude of indirect, second-order transition, involving an intermediate (the third) site. The interference exists and can be important for hops involving any number of phonons. However, for one-phonon hops it leads only to the interference mechanism of magnetoresistance~\cite{NSS,SS,our-int}. When all the relevant hops are included in the theory each triad of localization sites starts to act as a source of the Hall current.

 The study~\cite{Holstein} was focused on the a.c. current in a system where the number of electrons is small, compared to the number of localized states. The a.c. current can be described in terms of  averaging the Hall current over all the Hall sources. The problem of d.c. current is more complex and involves the generated Hall current distribution over the network of Miller-Abrahams resistors~\cite{Mil-Abr,ES}. The most conventional approach to this problem is the percolation theory that treats the system in the limit of strong disorder.

 There are two possible reasons for having strong disorder in a system with hopping transport. The first reason, the position disorder, is related to the random positions of localization sites and is controlled by the dimensionless parameter $n^{1/d}a$, where $n$ is the concentration of sites, $a$ is the localization radius and $d$ is the system dimension. When this disorder is dominant, the system is in  Nearest Neighbor Hopping (NNH) regime. The temperature dependence of conductivity in this regime follows the Arrhenius law. Although the NNH regime can be realized in the experiment, it is not always easy to distinguish it from the transport due to the carrier activation into the conduction band. The percolation theory was applied to the Hall effect in the NNH regime in studies~\cite{bryksin,FP78,Galp}. It was shown that the Hall current is determined essentially by the rare optimal triads of the sites which form a junction for the percolation
 paths.  The resulting Hall mobility exponentially decreases in the limit of strong disorder $n^{1/d}a \rightarrow 0$.

Another reason of disorder is the random distribution of the localized state energies. The width of this distribution $\Delta \varepsilon$ should be compared to temperature $T$. The control parameter of the energy disorder is $\Delta \varepsilon/T$. When this disorder is sufficiently strong, the system is in the Variable Range Hopping (VRH) regime. It can be identified in the experiment due to the unique temperature dependence of conductivity, the Mott law \cite{Mott-orig,Ambeo} or the Efros-Shklovskii law \cite{ES} in the systems, where the Coulomb gap is essential. The percolation theory for the Hall effect in the VRH regime was discussed in Refs. \cite{bryksin,gru81,FP81,Fri82,Galp,Burkov}. Also the similar theories were developed for the anomalous Hall effect \cite{Burkov,Ano2}. Although the approach used in these studies was more or less the same, the results are surprisingly different.
 In Refs. \cite{gru81,Burkov} the Hall mobility $\mu_H$ is predicted to have the exponential dependence on temperature $\mu_H \propto \exp(-(T_0'/T)^{1/4})$, with $T_0'$ smaller than  $T_0$ in the Mott law \cite{gru81}. In Refs. \cite{Fri82,Galp} the dependence follows the power law $\mu_H \propto T^{\gamma}$. The power law dependence appeared from the contribution of rare optimal triads of localization sites. These triads consist of three sites close to each other, but with the energies that lead to Miller-Abrahams resistances between the sites of the triad similar to the critical resistance of the percolation cluster. To be effective, such a triad should be positioned in the intersection of the percolation paths \cite{Mott1}, as in the case of NNH conductivity. The characteristic correlation distance $L_h$ between such junctions has been evaluated in Ref. \cite{Skal} and turns out to be much larger than the hopping network correlation length.  As a result, strong mesoscopic effects are expected even in relatively large samples \cite{Galp}.

The drawback of the percolation theory is that it is rigorous only in the analytical limit of a small temperature and localization radius. It becomes especially important when the result of the percolation theory is governed by the extremely rare objects such as the discussed optimal triads. It is not always clear at what temperature these rare objects start to dominate over other triads that are not that effective as Hall current sources, but are significantly more numerous. To understand this, the percolation theory can be compared to the numerical simulations based on the direct solution of Kirchhoff equations. Such attempts were made in studies~\cite{num1,num2}. These simulations support expression $\mu_H \propto \exp(-(T_{0}'/T)^{1/4})$. However, the computations in Refs.~\cite{num1,num2} were made with only one disorder realization in a 3D cubic sample with a size $\sim 13\times13\times13$ sites. It is not clear if this sample can represent a macroscopic disordered system. Also these results cannot be directly applied to 2D systems which are the main focus of our study.

The most important problems in the experimental study of hopping Hall effect are the small Hall resistance value
and the fact that the Hall effect is masked by
magnetoresistance. Nevertheless, a few experiments on the Hall conductivity in the hopping regime  were made in Refs.~\cite{Koon, Essalen, Zhang, Avdonin, Kajikawa, Amitay}.
Amitay and Pollak \cite{Amitay} attempted to measure the impurity-hopping Hall effect in germanium and silicon at a sufficiently low temperature and impurity concentration so  that any
Hall effect from carriers in delocalized states would be negligible. The authors have not succeeded in the observation of Hall effect  even though the sensitivity of their measuring system seemed to be sufficient to
detect it. The negative result was also obtained in Ref.~\cite{Klein},  measured in the system in the deep localization regime. The  Hall effect due to hopping conductivity was not detected, and it set
an upper bound on the Hall conductivity of 1.7$\times10^{-13}$ Ohm$\cdot$ cm$^{-1}$
for the given conditions. Most of the experimental observations of Hall effect in the dielectric regime were obtained near the metal-insulation transition (MIT) in the 3D case~\cite{Koon}, where a small range of $\sigma_{xx}(T)$ dependence was observed and the absolute value of  $\sigma$ was rather large. It means that the system can not be  in a strong localization regime.

The recent advances in the technology allowed us to grow the arrays of Si/Ge quantum dots (QDs) that display the VRH conductivity in the Coulomb gap regime~\cite{Yak00}. The interesting property of these arrays is that the localization radius near the Fermi level  is much larger than that for impurities in semiconductors. It is comparable or  larger than the QD size and interdot distance, and can be controlled by changing the filling factor. The possibility to change the structural parameters of QDs allows the novel way to control the disorder not possible in ordinary doped semiconductors.  Recently, we obtained the first experimental results of the Hall effect  in this system~\cite{Ste17}. To understand the obtained experimental data we need to compare our results to the theory. However, the theoretical results themselves are not well-established; therefore, we have to reconsider the theory of the Hall effect in the systems with hopping transport before the comparison can be made.

Modern computation potentials allow us to significantly improve the numerical approach~\cite{num1,num2}. As a result, we can verify the results of percolation theory and understand the correct dependence of the Hall mobility on the system parameters for the moderately low temperatures reachable in the experiment. We focus our study on the comparison of the percolation arguments with numeric simulations and with our recent experimental results. Also we restrict ourselves to the 2D case that was not treated numerically in Refs.~\cite{num1,num2} and is relevant for our experiments.

The paper is organized as follows. In Sec.~\ref{sec_gen} we derive the general equations in the form that allows both analytical and numeric treatments. In Sec.~\ref{sec_disorder} we consider our equations in strongly disordered systems in NNH, Mott law and Efros-Shklovskii law  VRH regimes and compare the  percolation arguments with the results of numeric simulation. In Sec.~\ref{sec_exp} we compare our theory with the experimental results. In Sec.~\ref{sec_discus} we provide a general discussion of the obtained results.

\section{General equations}
\label{sec_gen}

In this section we extend the approach \cite{Galp} to the description of the hopping Hall effect to include the triads of sites with arbitrary occupation numbers. First, we consider the interaction of localized electrons with phonons in the density matrix formalism. We derive the rate equations that describe the ordinary two-site one-phonon hops controling the conductivity and three-site two-phonon hops responsible for the Hall effect. These equations are then linearized to study the linear response to electric and magnetic fields. It leads to the system (\ref{lin0}-\ref{lin-G}) of the modified Kirchhoff equations, that is a useful starting point for both numerical simulations and percolation treatment.

In our study we adopt the model when the electrons are localized on point-like sites. The sites have random energies $\varepsilon_i$ that are larger than the overlap integrals $t_{ij}$. This model is conventional to the hopping transport. In real systems, for example in quantum dot arrays with hopping conduction, the physics can be more complex. It can include the finite size of a quantum dot and the importance of states of intermediate dots for the long-range hopping. However, the point-like site model is known to be a good starting point to study the hopping transport. It was assumed in most of the previous studies in the field \cite{bryksin,FP78,gru81,FP81,Fri82,Galp,Burkov,num1,num2}. Therefore, we think that it is instructive to achive a reliable understanding of the hopping Hall effect in the point-like site model before starting to consider peculiarities of complex systems. We also do not consider the electron spin to make our model as simple as possible.

The starting point of our consideration is the Hamiltonian of a system with hopping transport after the polaron transformation \cite{Bryksin-Book}
\begin{equation}\label{H1}
H = \sum_i \varepsilon_i n_i + \widehat{T} + H_{ph}, \quad \widehat{T} = \sum_{ij} t_{ij} \Phi_{ij} a_i^+ a_j.
\end{equation}
Here $\varepsilon_i$ is the on-site energy, $n_i = a_i^+ a_i$ is the operator of electron density on site $i$. The overlap integral $t_{ij}$ in the magnetic field can be expressed as
 $t_{ij} = t_0 \exp(-r_{ij}/a)\exp(i\frac{e}{2c} {\bf B} \cdot [{\bf r}_i\times {\bf r}_j])$, where ${\bf B}$ is the applied magnetic field. ${\bf r}_i$ is the position of site $i$, $r_{ij} = |{\bf r}_i - {\bf r}_j|$ is the distance between sites $i$ and $j$, $a$ is the localization length. In ordinary doped semiconductors far from the metal-insulator transition $a$ corresponds to the localization radius of a single impurity state. In semiconductors close to the metal-insulator transition and in more complex systems, $a$ can be strongly re-normalized by co-tunneling processes. It is especially important for granular metals and quantum dot arrays \cite{Shk-QD}. This re-normalization can easily make $a$ larger than the inter-site distance \cite{Shk-QD}.  $H_{ph} = \sum_{\bf q} \omega_q b_q^+ b_q$ is the phonon Hamiltonian.   There is no on-site electron-phonon interaction due to the polaron  transformation, however, the interaction is included in the transition elements that are proportional to
\begin{multline}\label{Phi-ij}
\Phi_{ij} =  \exp\Biggl\{ \sum_{\bf q}  b_{\bf q}^+ (u_j^*({\bf q}) - u_i^*({\bf q}))  - \\
   b_{\bf q}(u_j({\bf q}) - u_i({\bf q}))
  \Biggr\}.
\end{multline}
Here $u_i({\bf q}) = (2{\cal N})^{1/2}\gamma({\bf q}) \exp(-i{\bf q} {\bf r}_i)$,
${\cal N}$ is the number of atoms in the lattice and
$\gamma({\bf q})$ describes the electron-phonon interaction \cite{Bryksin-Book}.

The electron-electron Coulomb interaction does not contribute explicitly to the Hamiltonian (\ref{H1}). We assume that it can be added to the energy $\varepsilon_i \rightarrow \varepsilon_i + \sum_j e^2 n_j/r_{ij}$ to include the effects of the Coulomb gap or the Coulomb glass, however the Coulomb energy does not interfere with the hopping process itself.
In the theory of hopping transport it is assumed that the hopping rates are small compared to  frequencies $|\varepsilon_i - \varepsilon_j|/\hbar$. Accordingly, we expand the electron density matrix $\widehat{\rho}$ over small hopping rates. With the Hartree decoupling~\cite{Bryksin-Book} we assume that the zero-order density matrix can be expanded as a product
\begin{equation}\label{rho0}
\widehat{\rho}^{(0)} = \prod_i \widehat{\rho}^{(0)}_i, \quad \widehat{\rho}^{(0)}_i = f_i|1\rangle \langle 1 | + (1-f_i ) |0\rangle \langle 0 |.
\end{equation}
 Here $\widehat{\rho}^{(0)}_i$ is the density matrix on  site $i$. It  corresponds to some probability $f_i$ for site $i$ to have an electron. In equilibrium, $f_i$ is the Fermi function $f_i = 1/(e^{(\varepsilon_i-\mu)/T} + 1)$. The zero-order density matrix corresponds to the situation when the electrons rest on their localization sites.

The dynamics of  density matrix $\widehat{\rho}$ can be described with the series of perturbation equation
\begin{equation}\label{rho-p}
\widehat{\rho}^{(n+1)}(t) = -i \int_{-\infty}^t [\widehat{T},\widehat{\rho}^{(n)}(t')]dt'.
\end{equation}
The structure of matrix $\widehat{\rho}^{(n)}$ that appears due to the dynamics (\ref{rho-p}) is more complex than the structure of $\widehat{\rho}^{(0)}$. The operator $\widehat{T}$ mixes different localization states and, therefore, $\widehat{\rho}^{(n)}$ includes the elements that correspond to electron transition from one localization site to another.

We apply the reduction procedure to separate the small transition elements from the actual hopping process that changes the filling numbers of the sites. To calculate the addition to the density matrix $\widehat{\rho}^{(0)}_i$ due to the hopping we take the trace over phonon states and all the other sites, i.e. $\delta \widehat{\rho}_i = {\rm Tr}_{{\rm ph}, j\ne i} (\widehat{\rho} - \widehat{\rho}^{(0)})$. Finally, to describe the electron dynamics as hopping we should assume that the dynamics of the reduced density matrices $\widehat{\rho}^{(0)}_i$ is much slower than the oscillations of perturbations to this matrix that occur at frequencies $\sim (\varepsilon_i - \varepsilon_j)/\hbar$. It allows us to substitute $\widehat{\rho}^{(0)}_i (t')$ with $\widehat{\rho}^{(0)}_i(t)$ in the expression (\ref{rho-p}).

The effect of ordinary two site one phonon hops is then expressed as a reduction of the second order density matrix $\widehat{\rho}^{(2)}$ where the phonon exponents in (\ref{Phi-ij}) are expanded up to the second order of phonon creation and annihilation operators. It leads to the conventional equations for the hopping rates. To consider the Hall effect, one should include three-site, two phonon hops that are described by the reduced third-order density matrices $\widehat{\rho}^{(3)}$, where the phonon exponents are expanded up to the fourth order over the phonon creation and annihilation operators.

With the assumptions mentioned above we obtain the hopping transport equations that include three-site two-phonon hops.
\begin{multline}\label{eq-f}
\frac{d f_i}{dt} = \\
 \sum_{j \ne i} f_j(1-f_i) \left( W_{ij} + \sum_{k \ne i,j} W_{ikj}^{(0)}(1-f_k) + W_{ikj}^{(1)}f_k \right) - \\
 f_i(1-f_j)\left( W_{ji} + \sum_{k \ne i,j} W_{jki}^{(0)}(1-f_k) + W_{jki}^{(1)}f_k \right).
\end{multline}
Here $W_{ij}$, $W_{ikj}^{(0)}$ and $W_{ikj}^{(1)}$ describe the rates of $j \rightarrow i$ hops and yield the contributions to $df_i/dt$ proportional to
$f_j(1-f_i)$. The rate $W_{ij}$ stands for the ordinary two-site hop. The rate $W_{ikj}^{(0)}$ stands for the two-phonon hop involving the intermediate site $k$ that is assumed to be free. Therefore, its contribution is proportional to $(1-f_k)$. In the similar way, $W_{ikj}^{(1)}$ describes the three-site hop involving the filled site $k$. The corresponding contribution to $df_i/dt$  includes the term $f_k$. For a given pair $ij$, the role of the intermediate site can be played by any  site of the system other than $i$ and $j$.

The rates  $W_{ij}$,  $W_{ikj}^{(0)}$ and $W_{ikj}^{(1)}$ are expressed as follows
\begin{equation}\label{Wij}
W_{ij} = \frac{1}{\tau_0} \exp(-2r_{ij}/a) {\cal N}(\varepsilon_j - \varepsilon_i),
\end{equation}
\begin{multline} \label{Wikj0}
W_{ikj}^{(0)} = \frac{1}{4} |t_{ij} t_{jk} t_{ki}| \frac{{\bf B} {\bf S}_{ikj}}{2\Phi_0} \times \\
\left( \frac{W_{kj} W_{ik}}{|t_{kj}^2 t_{ik}^2|}
+\frac{W_{ij} W_{kj}}{|t_{ij}^2 t_{kj}^2|}
+ \frac{W_{ij} W_{ki}}{|t_{ij}^2 t_{ik}^2|}
 \right),
\end{multline}
\begin{multline} \label{Wikj1}
W_{ikj}^{(1)} = -\frac{1}{4} |t_{ij} t_{jk} t_{ki}| \frac{{\bf B} {\bf S}_{ikj}}{2\Phi_0} \times \\
\left( \frac{W_{kj} W_{ik}}{|t_{kj}^2 t_{ik}^2|}
+\frac{W_{ij} W_{jk}}{|t_{ij}^2 t_{kj}^2|}
+ \frac{W_{ij} W_{ik}}{|t_{ij}^2 t_{ik}^2|}
 \right).
\end{multline}
Here $\tau_0$ is the constant describing the characteristic (ordinary) hopping time between close neighbors without a large energy exponent. The similar time for three site hops is $t_0 \tau_0^2/\hbar$. It is assumed to be small, compared to $\tau_0$. ${\cal N}(\varepsilon_j - \varepsilon_i)$ is the effective probability to find a phonon for the hop. We consider ${\cal N}(\varepsilon_j - \varepsilon_i) = 1$ when $\varepsilon_j > \varepsilon_i$ and ${\cal N}(\varepsilon_j - \varepsilon_i) = \exp((\varepsilon_j -\varepsilon_i)/T)$ otherwise. ${\bf S}_{ikj}$ is the ``vector area'' of the triangle $ikj$, $\Phi_0$ is the flux quanta. We focus on small magnetic fields ${\bf B}{\bf S}_{ikj} \ll \Phi_0$. In the general case, the linear dependence on the magnetic field should be substituted with oscillating dependence $\sin({\bf B}{\bf S}_{ikj} / 2\Phi_0)$. The expressions (\ref{Wij} - \ref{Wikj1}) are derived from (\ref{rho-p}) in  appendix \ref{App1}. These expressions agree to \cite{Holstein,Galp} when the triangle $ikj$ is considered to have small occupation numbers $f_i$, $f_j$ and $f_k$. However the expressions (\ref{Wij} - \ref{Wikj1}) allow the description of the general case of the arbitrary occupation numbers.

 We discuss the system in the  Ohmic regime. It corresponds to the small perturbations of occupation probabilities $f_i$. This case allows using the linearized version of the general equations. We consider the dc current and stationary equations $d f_i/dt = 0$.
\begin{equation}\label{lin0}
\sum_j J_{ij} = 0,
\end{equation}
\begin{equation}\label{lin1}
J_{ij} = \frac{\varphi_i - \varphi_j}{R_{ij}} +
\sum_{k\ne i,j} \frac{e^2\varphi_k}{T} \frac{{\bf S}_{ikj} {\bf B}}{2\Phi_0} {\Gamma}_{ikj},
\end{equation}
\begin{equation}\label{R-MA}
R_{ij} = \frac{T}{e^2\Gamma_{ij}}, \quad \Gamma_{ij} = \frac{1}{\tau_0} \exp\left(-\frac{2r_{ij}}{a} - \frac{\varepsilon_{ij}}{T}\right).
\end{equation}
Here $R_{ij}$ is the Miller-Abrahams resistor between sites $i$ and $j$, which has an exponentially-broad distribution $R_{ij} \approx R_{0} \exp(\xi_{ij})$, $\xi_{ij} = 2r_{ij}/a + \varepsilon_{ij}/T$  in a material with a strong disorder.  $\varepsilon_{ij} = (|\varepsilon_i - \varepsilon_F| + |\varepsilon_j - \varepsilon_F| + |\varepsilon_i - \varepsilon_j|)/2$ is the energy term in the Miller-Abrahams resistance expression. $\varepsilon_F$ is the Fermi energy. $\varphi_i$ is the addition to the electrochemical potential of site $i$ due to the applied current.  $\Gamma_{ikj}$ is the rate of the three-site hop.  In a strongly disordered system it can be estimated as
\begin{multline}\label{lin-G}
{\Gamma}_{ikj} = \frac{1}{4t_0\tau_0^2} \exp\left(-\frac{r_{ij}+r_{ik} + r_{kj}}{a}\right)  \times \\
\left(
e^{(|\varepsilon_i-\varepsilon_F| - \varepsilon_{ij}-\varepsilon_{ik})/T} +
e^{(|\varepsilon_j-\varepsilon_F| - \varepsilon_{ij}-\varepsilon_{jk})/T} + \right. \\
\left.
e^{(|\varepsilon_k-\varepsilon_F| - \varepsilon_{ik}-\varepsilon_{jk})/T}
\right).
\end{multline}

In our expressions for the hopping rates $\Gamma_{ij}$ and $\Gamma_{ijk}$, we keep only the exponential terms in the dependencies on $r_{ij}$ and $\varepsilon_{ij}$. This approximation can be applied in strongly disordered systems when the power-law terms are small compared to the exponential ones.

\section{Hall current in systems with strong disorder}
\label{sec_disorder}

In this section we study the hopping Hall effect in systems with strong disorder. We apply two methods to this problem. The first one is the direct numerical solution of
the system  (\ref{lin0}-\ref{lin-G}) of modified Kirchhoff equations. The second is the analytical approach based on the percolation theory. In the frame of the analytical method, we evaluate the contribution of
different triads of sites to the Hall effect. Assuming that the effect is dominated by a small number of the so-called optimal triads, we derive the analytical expression for Hall mobility. The main purpose of this section is to compare the results of  numerical and percolation approaches. We provide this comparison for the three important cases: nearest neighbor hopping (Sec.~\ref{NNH}), variable range hopping with a constant density of states (Sec.~\ref{VRH-Mott}) and variable range hopping in the Coulomb gap (Sec.~\ref{VRH-ES}).

In the numerical simulation we consider a square 2D numerical sample with size $L$ and $N=L^2$ localization sites with random positions. The positions are not correlated. Each site $i$ is ascribed with some energy $\varepsilon_i$. The localization distance $a$,  temperature $T$ and the distribution of the site energies control the degree of disorder and the hopping regime (NNH or VRH). The boundary conditions are periodical. It means that our simulation represents the infinite system composed of $L\times L$ supercells.

The localization sites distribution  determines the system of linear equations (\ref{lin0}-\ref{lin-G}). The system is solved numerically without the magnetic field and with small magnetic field $B$ when the corrections to the currents are linear with respect to the field $B$. We find the normal current $J$ at $B=0$ directed along the electric field that gives us conductivity $\sigma$ and the current $J_{Hall}\propto B$ that is perpendicular to the electric field. The Hall mobility $\mu_H$ is proportional to the ratio of these currents $\mu_H = J_{Hall}/JB$. The results are then averaged over disordered configurations.
With numeric simulations we are able to find the $\mu_H$ dependence   on localization radius $a$ and temperature. The absolute value of $\mu_H$ is governed by the parameter $t_0 \tau_0/\hbar$ that is not discussed in the present study.

Our analytical treatment of equations (\ref{lin1}) relies on the exponentially-broad distribution of coefficients  ${\Gamma}_{ikj}$ in a strongly disordered system. As soon as we discuss the linear effect over magnetic field ${\bf B}$,  we can consider  potentials $\varphi_k$ in the last term in (\ref{lin1}) to be independent from the magnetic field. Values $\varphi_k$ are proportional to the applied electric field and are determined by the disorder configuration. In this case, each triad of sites $ikj$ acts as an independent source of the Hall current. The total Hall current $J_{hall}$ flowing trough some cross-section of the sample (perpendicular to the electric field direction) can be expressed as the sum of contributions $J_{hall}^{(ikj)}$ related to triads $ikj$.
\begin{equation}\label{Jhall-sum}
J_{hall} = \sum_{ikj}J_{hall}^{(ikj)}.
\end{equation}
The formal definition of contribution $J_{hall}^{(ikj)}$ is as follows. We consider the system where the magnetic flux  exists only in the triad $ikj$, but the other properties are the same as in initial system. The Hall current in this modified system is equal to $J_{hall}^{(ikj)}$.
Note that contributions $J_{hall}^{(ikj)}$ depend not only on the properties of the triad itself, but also on its position, with respect to the percolation cluster. Contributions $J_{hall}^{(ikj)}$ have an exponentially-broad distribution and their sum is assumed to be controlled by a small number of the largest $J_{hall}^{(ikj)}$. The idea of the percolation analysis is to identify these largest contributions and neglect all other ones that are exponentially small compared to ${\rm max}(J_{hall}^{(ikj)})$.

\begin{figure}[htbp]
    \centering
        \includegraphics[width=0.4\textwidth]{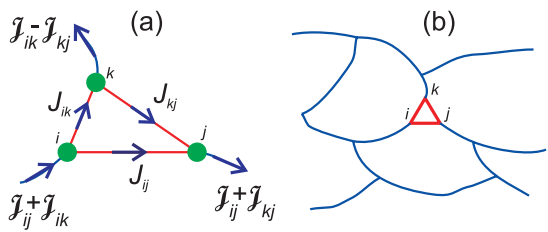}
        \caption{(a) The currents in triangle $ikj$, (b) the optimal position of the triangle $ikj$ in the percolation cluster.}
    \label{fig:tri}
\end{figure}

Let us start from the properties of the triad  itself. The currents that flow through the triad can be described as three currents ${\cal J}_{ij}$, ${\cal J}_{ik}$ and ${\cal J}_{kj}$ (Fig.~\ref{fig:tri} (a)). The effects of the currents are additive. Let us consider one of them, ${\cal J}_{ij}$ and assume ${\cal J}_{ik} = {\cal J}_{kj} = 0$. Current ${\cal J}_{ij}$ enters  site $i$, then it is divided between $J_{ij}$ and $J_{ik}=J_{kj}$. Then it flows through the site $j$ out of the triad.

According to equation (\ref{lin1}), the site $k$ in the magnetic field acquires a perturbation to its potential that is proportional to ${\cal J}_{ij}$
\begin{equation}
\delta \varphi_k = {\cal L}_{kij} {\cal J}_{ij}, \quad {\cal L}_{kij} = \frac{R_{ij} R_{ik} R_{kj}}{R_{ij}+R_{ik} + R_{kj}} \frac{{\bf S}_{ikj} {\bf B}}{T} \Gamma_{ikj}.
\end{equation}
Here ${\cal L}_{kij} = {\cal L}_{ijk} = {\cal L}_{jki}$. Value ${\cal L}_{kij} {\cal J}_{ij}$ can be considered as a source of Hall voltage.  ${\cal L}_{kij}$ values have an exponentially-broad distribution
\begin{equation}
{\cal L}_{kij} \approx {\cal L}_0 \frac{{\bf S}_{ikj} {\bf B}}{T} \exp(\xi_{ikj}),
\end{equation}
\begin{multline} \label{xi-ikj}
\xi_{ikj} = \xi_{ij}+\xi_{ik} + \xi_{kj} - \\  {\rm max}(\xi_{ij},\xi_{ik},\xi_{kj}) -
 \frac{r_{ij} + r_{jk} + r_{ik}}{a} - \\
 {\rm min}\Bigl(
\frac{\varepsilon_{ij} + \varepsilon_{ik} - |\varepsilon_i|}{T},
\frac{\varepsilon_{ij} + \varepsilon_{jk} - |\varepsilon_j|}{T},
\frac{\varepsilon_{ik} + \varepsilon_{jk} - |\varepsilon_k|}{T}
\Bigr).
\end{multline}
Current ${\cal J}_{ij}$ is of the order of the ``percolation current'' $J_{perc} = L_{cor} j$ when  resistor $R_{ij}$ is included into the percolation cluster. Here $L_{cor}$ is the correlation length of the percolation cluster, $j$ is the macroscopic current density. Therefore, reasonably large currents are possible when  resistor $R_{ij}$ is not larger than the critical resistance of  percolation network $R_{c}$. Otherwise current ${\cal J}_{ij}$ becomes small ${\cal J}_{ij} \sim (R_{c}/R_{ij}) L_{cor}j$ for $R_{ij} > R_{c}$. In this case, triangle $ikj$ cannot be an effective Hall current source.

 The contribution $J_{hall}^{(ikj)}$ of triad $ikj$ depends on its position  in the Miller-Abrahams resistor network. If it is shunted by resistances $R_{nm} \ll R_{ij}, R_{ik}, R_{kj}$, its contribution to the Hall current is small. The most effective sources are composed of resistors $R_{ij}\sim R_{ik} \sim R_{kj} \sim R_{c}$  and are positioned in the junction of three branches of the percolative cluster (Fig.~\ref{fig:tri} (b)). Hall mobility $\mu_{hall}$ can be, thus, estimated as
 \begin{equation}
\mu_{hall} \approx p_{\triangle}{\cal L}_{kij}^{(max)} /R_{perc},
\end{equation}
where ${\cal L}_{kij}^{(max)}$ is the maximum possible value of ${\cal L}_{kij}$ for the triangles that allow the percolation current ${\cal J}_{ij} \sim J_{perc}$. $p_{\triangle}$ is the probability of three branches of the percolative cluster to be connected by the optimal triad. With the exponential precision, the mobility can be estimated as $\mu_{hall} \propto \exp({\rm max}(\xi_{ikj}) - \xi_{c})$, where $\xi_c$ is the critical exponent of the percolation theory.

\subsection{Nearest neighbor hopping}
\label{NNH}

In the systems with the nearest neighbor hopping conduction the distribution of site energies is not broad compared to the temperature. All the disorder comes from the random positions of localization sites. It is controlled by parameter $na^2$. We focus on the strong disorder case $na^2 \ll 1$. Therefore we neglect the energy terms in (\ref{xi-ikj}) and get
\begin{equation}
\xi_{ikj} = \frac{r_{ij} + r_{ik} + r_{kj}}{a} - 2 \frac{{\rm max}(r_{ij},r_{jk},r_{ik})}{a}.
\end{equation}
The maximum possible value is $\xi_{ikj} = r_{perc}/a$ where $r_{perc} = \sqrt{4/\pi} n^{-1/2}$ is the percolation distance of the random site percolation problem \cite{ES}. The critical resistance in this system is equal to $R_{c} = R_0 \exp(2r_{perc}/a)$. The Hall mobility follows the law
\begin{equation}\label{muHnn}
\mu_{hall} \propto \left( \frac{a}{r_{perc}}\right)^{\gamma_{nn}} \exp\left( - \frac{r_{perc}}{a}\right).
\end{equation}
Here $\gamma_{nn}$ describe the power law dependence of $p_{\triangle}$ on the localization radius. The conductivity depends on $r_{perc}$ as $\sigma = \sigma_0\exp(-2r_{perc}/a)$.

\begin{figure}[htbp]
    \centering
        \includegraphics[width=0.4\textwidth]{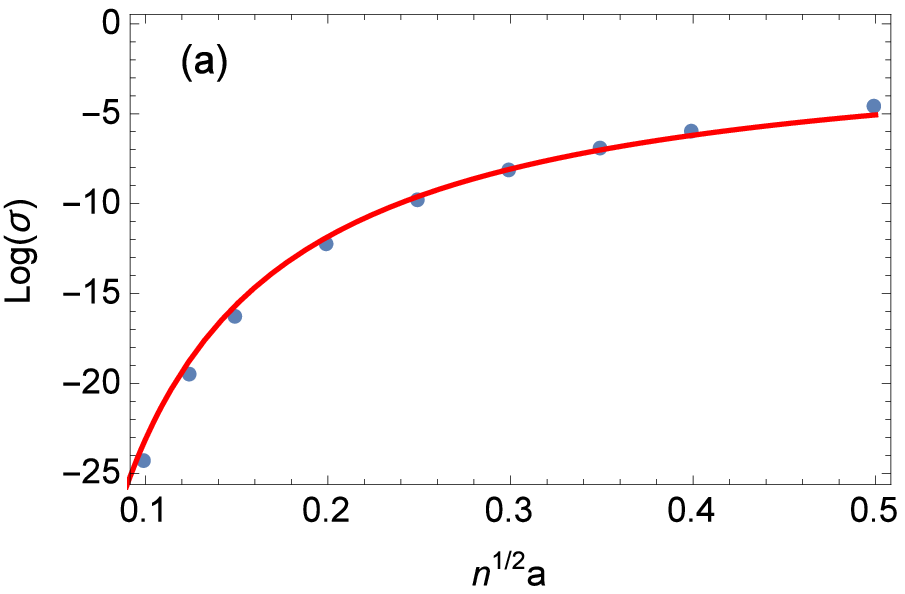}
        \includegraphics[width=0.4\textwidth]{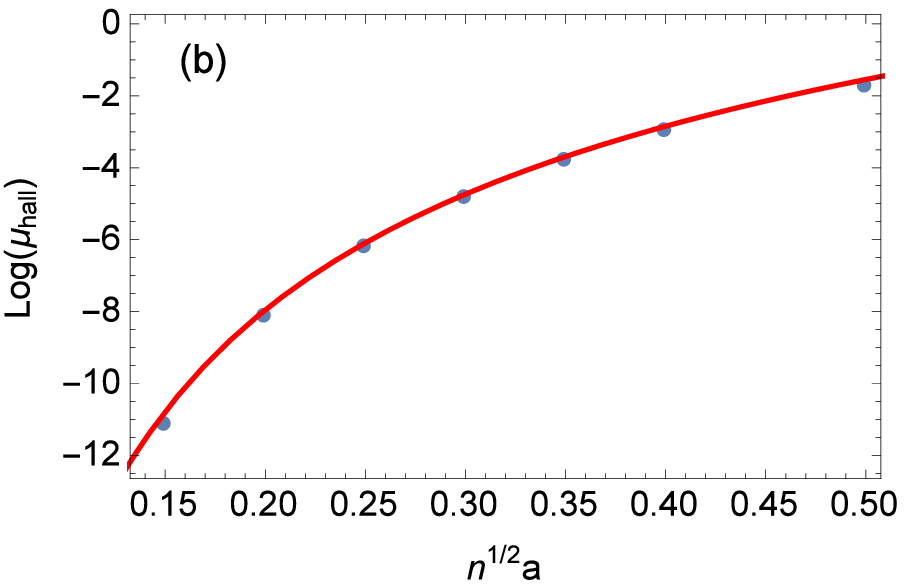}
        \caption{The results of the numerical simulation of neighbor hopping transport. (a) the simulated conductivity (blue dots) compared to the law    $\sigma = \sigma_0 \exp(-2r_{perc}/a)$ (red curve);
        (b) the simulated  Hall mobility (blue dots) compared to eq. (\ref{muHnn}) with $\gamma_{nn}=3.3$ (red curve).  }
    \label{fig:nei}
\end{figure}

For the neighbor hopping regime in the numerical simulation, we consider numerical samples without random energies of localization sites. The site concentration is equal to unity and the positions of sites are random with Poisson distribution. Localization radius $a$ controls the disorder. The equations (\ref{lin1}-\ref{lin-G}) are solved numerically for the system to find the dependence of conductivity $\sigma$ and Hall mobility $\mu_H$ on the disorder parameter $n^{1/2}a$. The results are shown in  Fig.~\ref{fig:nei}. They are in a good agreement with the analytical predictions for $\gamma_{nn} = 3.3$.

\subsection{Variable range hopping}
\label{VRH-Mott}

In the variable range hopping (Mott law) regime  the site energies $\varepsilon_i$ are large compared to temperature. The density of states has no peculiarities at the Fermi level and can be described by a constant $g(\varepsilon) \approx g(\varepsilon_F)$. The exponents $\xi_{ij}$ in the resistors contain both the coordinate contribution $2r_{ij}/a$ and the energy contribution $\varepsilon_{ij}/T$. Therefore the large exponent $\xi_c$ of the critical resistor can reflect the long inter-site distance between sites $i$ and $j$ or the large energies of sites.

The long distances between sites in the Hall source triad $ikj$ yield exponentially small (as a function of disorder) Hall mobility $\mu_H$ as it was shown in the previous section. The situation with large energies is different. Let us imagine the triangle $ikj$ composed of the three close sites with energies $\varepsilon \approx \xi_c T$. The equation (\ref{xi-ikj}) yields the Hall source exponent of this triangle $\xi_{ikj} = \xi_c$. It can be shown that it is the maximum possible value for $\xi_{ikj}$. It is equal to the critical exponent of the conductivity. It means that if the probability $p_{\triangle}$ of finding three branches of the percolation cluster connected with such an optimal triangle has a power law dependence on temperature, the dependence $\mu_{hall}(T)$ should follow the same power law. The power law for the $p_{\triangle}(T)$ dependence is a natural assumption because the distribution of site energies and positions in the percolation cluster follow power laws. However, it will be discussed in some details in Sec.~\ref{sec_discus} and the counter-arguments for the power law will be provided. Now we want to note that even if the dependence $p_{\triangle}(T)$ follows the power law, the dominance of the optimal triangles over the Hall effect is proved only in $T\rightarrow 0$ limit. At finite temperatures it should be verified with a numerical simulation.

For the Mott law regime in the numerical simulation we add random energies to the sites of our numerical samples. The distribution of energies has the constant density $g(\varepsilon) = n/\Delta \varepsilon$ in some energy range $-\Delta \varepsilon/2 < \varepsilon < \Delta \varepsilon/2$. The Fermi level is assumed to be equal to zero. The parameter $\Delta \varepsilon/T$ controls the energy disorder.  The analytical expression for the conductivity in this regime, the Mott law, is
\begin{equation}\label{Mott}
\sigma = \sigma_0 \exp \left[-(T_{0}/T)^{1/3}\right].
\end{equation}
Here $T_{0} = \beta_{2D}/g(\varepsilon_F)a^2$, $\beta_{2D} = 13.8$~\cite{ES}, $g(\varepsilon_F)$ is the density of states at  Fermi level $\varepsilon_F$. It follows from the percolation theory and is valid in the limit of strong disorder $T_{0}/T \gg 1$. The pre-factor $\sigma_0$ can have a power-law dependence on the system parameters that can be derived from the system dimension and the power-law part of the $R_{ij}$ dependence  on the system parameters~\cite{ES}. In our case of a 2D system with $R_{ij} \propto T$, $\sigma_0$ is proportional to $1/T$ and is independent of the localization radius.

\begin{figure}[htbp]
    \centering
        \includegraphics[width=0.45\textwidth]{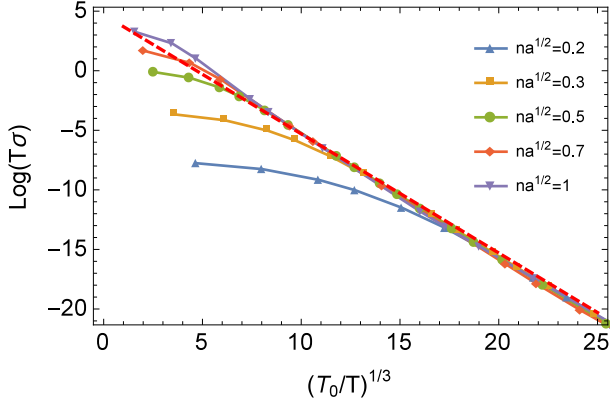}
        \caption{The simulated temperature dependence of conductivity in the Mott law VRH regime for  values  $n^{1/2}a=0.2$, $0.3$, $0.5$, $0.7$ and $1$. The Mott law is shown with a red dashed line.  }
    \label{fig:mott-sig}
\end{figure}

Therefore, in the VRH regime the product $\sigma T$ should depend only on $\xi_c = (T_0/T)^{1/3}$. The point, when the dependences of $\sigma T$ on critical exponent $\xi_c$ converge to the single curve, can be considered as a condition for the VRH conductivity.  In Fig.~\ref{fig:mott-sig}(a) we show the simulated dependence of $\ln(T\sigma)$ on $(T_{Mott}/T)^{1/3}$ for different values of the position disorder parameter $n^{1/2}a$. It is compared to the Mott law (red dashed line). It seems that the agreement starts from a relatively small $\xi_c = (T_{Mott}/T)^{1/3} \approx 5$ for the weak position disorder $n^{1/2}a  \lesssim 0.5$. For a stronger position disorder the VRH regime starts from larger $\xi_c$.

\begin{figure}[htbp]
    \centering
        \includegraphics[width=0.45\textwidth]{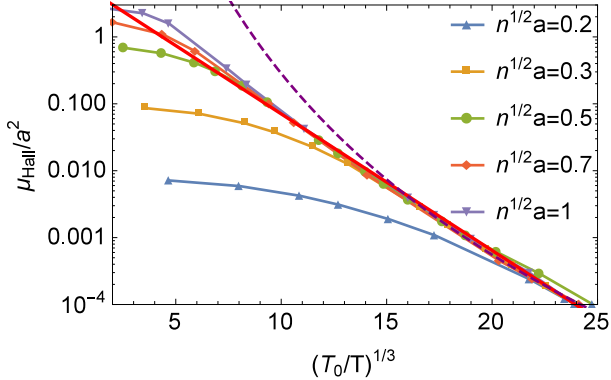}
        \caption{The numerical results for the Hall mobility. The results are compared to eq. (\ref{muH-mott}) (red straight line) and the power law dependence (purple dashed line). }
    \label{fig:mott-sc}
\end{figure}

Now let us discuss the dependencies of Hall mobility on the localization distance and temperature. In the VRH regime these dependencies are related. The transport properties that are determined by dimensionless distances $r_{ij}/a$ and energies $\varepsilon_i/T$ can depend only on the combined parameter $\xi_c = (\beta_{2D}/g(\varepsilon_F)a^2T)^{1/3}$ because only the states close to the Fermi energy are important in VRH. It follows from the scaling arguments. The change of the temperature and localization distance $T \rightarrow \widetilde{\beta} T$, $a \rightarrow \widetilde{\alpha} a$ is equivalent to the change of  site density $n \rightarrow \widetilde{\alpha}^2 n$ and all  site energies $\varepsilon_i \rightarrow \widetilde{\beta}^{-1} \varepsilon_i$. When $\widetilde{\alpha}^2\widetilde{\beta}=1$ (it is the condition for $\xi_c = const$) the density of states at the Fermi level $g(\varepsilon_F)$ stays constant and the  $g$ dependence on energy can be neglected in the Mott law regime.

However, the Hall sources are proportional to $B S_{ikj}/\Phi_0$. This value is not controlled by the $r_{ij}/a$ ratio but with the squared intersite distance compared to $\Phi_0/B$. The discussed scaling procedure will modify these terms $B S_{ikj}/\Phi_0 \rightarrow \widetilde{\alpha}^2 B S_{ikj}/\Phi_0$. Therefore in the VRH regime $\mu_{hall}$ should depend on the system parameters as follows.
\begin{equation}\label{scaling-norm}
\mu_{hall} = \frac{a^2}{\Phi_0} f(\xi_c).
\end{equation}
Here $f(\xi_c)$ is a function of  critical exponent $\xi_c$. The assumption that the Hall effect is controlled by the optimal triangles yield $f(\xi_c) = const\cdot \xi_c^{\gamma_{Mott}}$ with some power $\gamma_{Mott}$.

Our numerical results for Hall mobility are shown in Fig.~\ref{fig:mott-sc}. The dependencies of  $\mu_H/a^2$ on $(T_0/T)^{1/3}$ for different $n^{1/2}a$ converge to a single curve at sufficiently small temperatures. It indicates the applicability of the law (\ref{scaling-norm}). The curve is compared to two laws that were discussed in the previous studies. The red straight line corresponds to the exponential dependence
\begin{equation}\label{muH-mott}
\mu_{hall} \propto a^2\exp(-\alpha_{mott} (T_0/T)^{1/3}).
 \end{equation}
 The numerical estimate is $\alpha_{mott} = 0.47$. The purple dashed line corresponds to the power law dependence $\mu_{hall}/a^2 \propto (T_0/T)^{\gamma_{mott}/3}$. The agreement with the exponential dependence is better. However, at small temperatures, when the curves for small $n^{1/2}a$ converge to the universal dependence, the result can, in principle, be described with the power law, at least for the considered values of $T_0/T$.

To independently test the optimal triangle assumption, we discuss the average area of the triangles responsible for the Hall effect. The percolation theory predicts that the effect is controlled by the following triangles. At a high temperature and strong position disorder ${n^{1/2}a \ll 1}$ (NNH regime), the optimal triangle is the equilateral triangle with side  $r_{perc}$. Its area is equal to $(\sqrt{3}/4)r_{perc}^2 \approx 0.62 n^{-1}$. At a small temperature in the VRH regime, the area of the optimal triangle can be estimated as $a^2$. Note that it is smaller than $(\sqrt{3}/4)r_{perc}^2$.

The linear nature of Kirchhoff equations allows us to access the area of the optimal triangle in the numerical simulation. The equation (\ref{Jhall-sum}) states that the Hall current can be described as a sum of the contribution related to each triangle. Although the contributions themselves cannot be easily separated in the final results of simulation, we can make the following numerical experiment. We artificially multiply the value ${\cal L}_{ikj}$ for each Hall source to the absolute value of the area of the corresponding triangle $|S_{ikj}|$. Then we recalculate the Hall current and obtain its new value $J_{hall}^{(mod)}$. Finally, we divide the modified Hall current to the original Hall current and obtain the area $\langle S \rangle_{hall}$.
\begin{equation}
\langle S \rangle_{hall} = \frac{J_{hall}^{(mod)}}{J_{hall}} = \frac{\sum_{ikj}|S_{ikj}|J_{hall}^{(ikj)}}{\sum_{ikj}J_{hall}^{(ikj)}}.
\end{equation}
It is the area of the triangle averaged with  weight $J_{hall}^{(ikj)}$, the contribution of the triangle to the total Hall current.

\begin{figure}[htbp]
    \centering
        \includegraphics[width=0.4\textwidth]{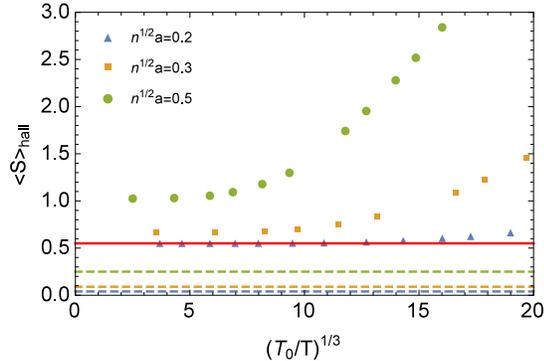}
        \caption{The averaged area of the triangle responsible for the Hall effect. The red straight line corresponds to value $(\sqrt{3}/4)r_{perc}^2$. Dashed lines correspond to the percolation theory predictions in the VRH regime.}
    \label{fig:AvArea}
\end{figure}

The numerical results for $\langle S \rangle_{hall}$ are shown in Fig.~\ref{fig:AvArea}. At high temperatures $\langle S \rangle_{hall}$ is slightly larger than $(\sqrt{3}/4)r_{perc}^2$ and tends to this value for a strong position disorder $n^{1/2}a \ll 1$. But, upon decreasing temperature, it increases instead of decreasing to its VRH-percolation value. It means that, at the considered parameters, the Hall current is dominated by the triangles that are much larger than the optimal triangle of the VRH percolation theory.

Another result of the percolation theory that we want to test with the numerical simulation is the prediction of very strong mesoscopic effects for the Hall current in VRH~\cite{Galp}. This prediction was based on the concept of optimal triangles. If the Hall effect is controlled by the rare triads of sites, the correlation length of the Hall effect should be proportional to the distance between these triads and be much larger than the correlation length of the percolation cluster. Therefore, it was suggested that even relatively large samples can show strong mesoscopic effects.

\begin{figure}[htbp]
    \centering
        \includegraphics[width=0.4\textwidth]{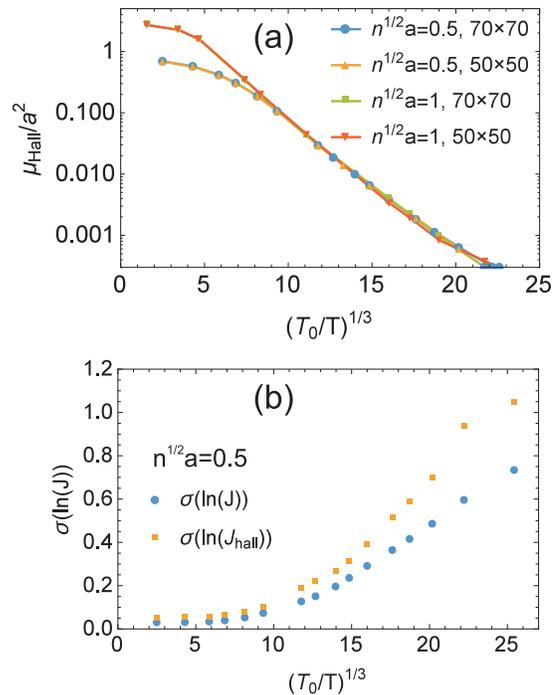}
        \caption{(a) the comparison of the Hall mobility in the numerical simulation with different sample sizes. (b) The standard deviation of the logarithm of normal and Hall currents.}
    \label{fig:mott-meso}
\end{figure}

It is also important to study mesoscopic effects to verify the applicability of our numerical results. The real samples are usually larger than our numerical samples. In Fig.~\ref{fig:mott-meso}(a) we compare the results for the numerical samples with size $70\times70$ considered in the rest of the present study with the results for smaller $50 \times 50$ numerical samples containing $2500$ sites. If the correlation length for the Hall effect is larger than our numerical samples one should expect significant difference in the Hall mobility calculated for different system size. However the results for $50 \times 50$ and $70\times 70$ systems are in a good agreement up to the smallest considered temperatures.

 In Fig.~\ref{fig:mott-meso}(b) we show the standard deviation of the logarithm of normal and Hall currents, $\sigma(\ln(J))$ and $\sigma(\ln(J_{hall}))$, correspondingly. When this deviation is small $\sigma(\ln(J)) \ll 1$, the fluctuations of the current are much smaller than the average current and the system is macroscopic. The opposite case $\sigma(\ln(J)) \gg 1$ corresponds to the exponentially-broad distribution of currents. This result is expected for the systems with the hopping transport that are smaller than the correlation length. The calculated standard deviation of $\ln(J_{hall})$ is slightly larger than the one for the normal current, but the difference is not dramatic. Both standard deviations are less than unity for the considered system parameters. It means that the mesoscopic effects for the Hall current should be only slightly larger than the ones for the ordinary conductivity and our numerical samples are larger than the correlation length for Hall effect at least for considered system parameters. Note that the size of our numerical samples ($4900$ sites) is still small, compared to the most real samples that are studied experimentally.

\subsection{Variable range hopping with the Coulomb gap}
\label{VRH-ES}

\begin{figure}[htbp]
    \centering
        \includegraphics[width=0.3\textwidth]{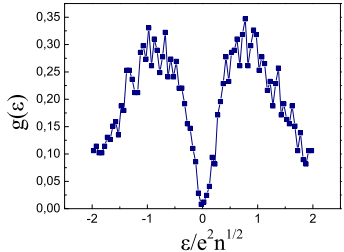}
        \caption{The density of site energies in one numerical sample obtained with the zero-temperature Monte Carlo simulation. }
    \label{fig:gE}
\end{figure}

Our experimental results are obtained in the samples that demonstrate the variable range hopping conductivity in the Efros-Shklovskii regime ${\sigma \propto \exp(-(T_{ES}/T)^{1/2})}$.
In this regime the Coulomb interaction between hopping electrons becomes essential. Strictly speaking the Miller Abrahams network cannot be rigorously derived for this case. Nevertheless, many important results for this regime are obtained by considering the resistor network where the Coulomb correlations were included as the Coulomb gap. In the present section we include the Coulomb gap to our system of Kirchhoff equations. We suggest, however, that our results can be dependent on the probability to find a triangle of critical resistors with special relations between distances and energies (as it is predicted by the percolation theory). The positions of sites in the Coulomb gap are correlated. Therefore, to keep these correlation in our system we do not simply ascribe each site a random energy with a distribution that includes the Coulomb gap. We follow a more complex procedure. We start with a numerical sample with random positions of the sites and consider a random half of them to be filled with electrons. Then we run the zero-temperature Monte-Carlo algorithm, i.e. we resolve all one-electron hops that decrease the total energy of the system including the electron-electron Coulomb repulsion. The details of this algorithm are given in Ref. \cite{Coulomb-ava}. It yields a meta-stable state of the system that naturally includes the Coulomb gap and the correlations in positions of sites with the energies close to the Fermi level. In the obtained state we find all one-electron energies and substitute them to the expressions for the Miller-Abrahams resistors and  Hall sources. The rest of the calculation is the same as in the regime of the Mott law. Our approach allows us to consider the ''static`` Coulomb correlations, however, it disregards the dynamic correlations, i.e. the modification of energies $\varepsilon_i$ due to the electron hops. Nevertheless, it allows the consideration of relatively large numerical samples deep in the VRH regime, which are not easy to access with the finite temperature Monte-Carlo algorithm (that includes all the dynamic Coulomb correlations).

\begin{figure}[htbp]
    \centering
        \includegraphics[width=0.4\textwidth]{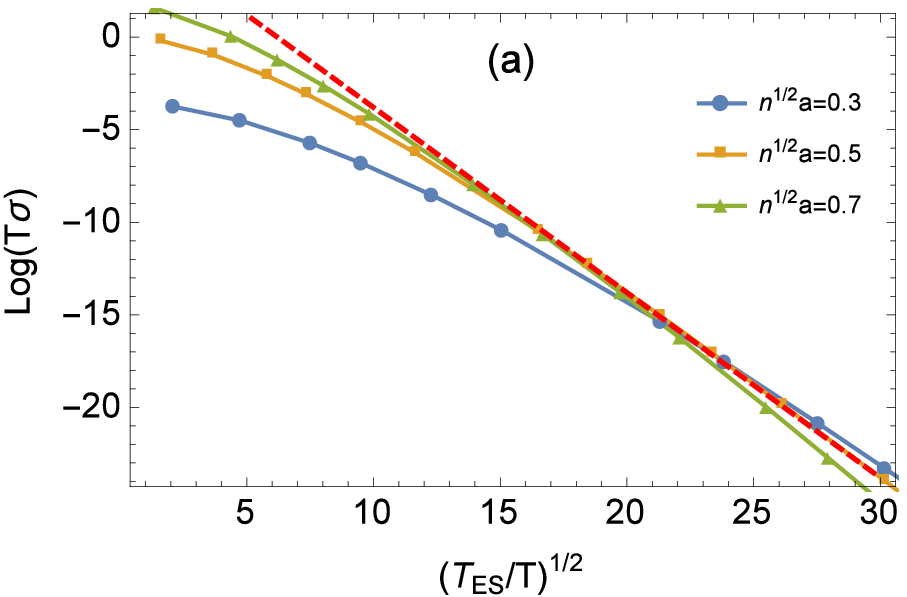}
        \includegraphics[width=0.4\textwidth]{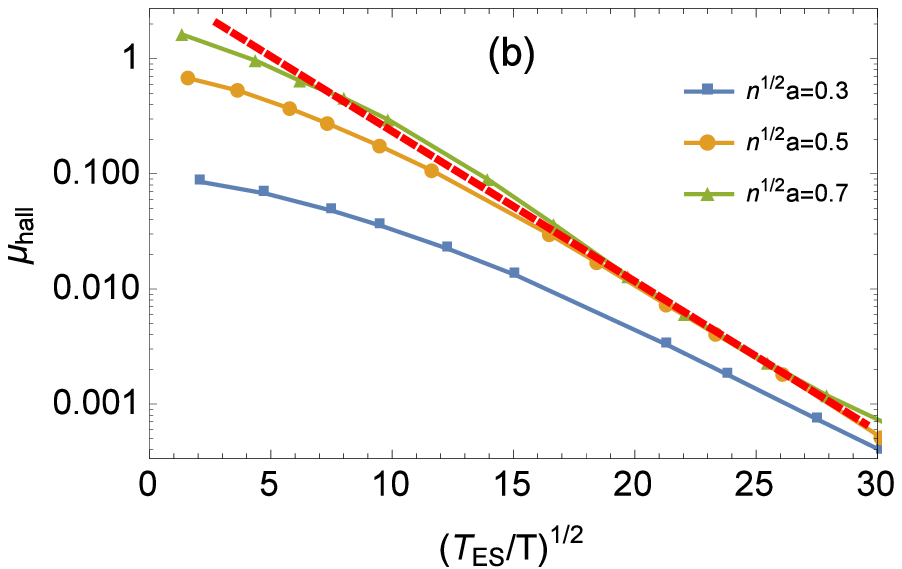}
        \caption{The results of numerical simulation of variable range hopping transport in the Efros-Shklovskii regime. (a) the conductivity compared to the Efros-Shklovskii law; (b) the Hall mobility compared to the expression (\ref{muH-ES}).   }
    \label{fig:ES}
\end{figure}

The distribution of the site energies obtained with our method in a single numerical sample is shown on Fig.~\ref{fig:gE}. It shows that, even in a single numerical sample, the Coulomb gap is well-defined.
 The results of simulation in the Efros-Shklovskii regime are shown in Fig.~\ref{fig:ES}. They are qualitatively similar to the results in the Mott law regime, however, the dependences of $T\sigma$ and $\mu_{hall}$ on $\xi_c$ converge to a universal curve more slowly than in the regime of the Mott law. We suggest that the reason for it is the double transition: from nearest neighbor hopping to VRH and from Mott VRH to Efros-Shklovskii VRH. The dependence of Hall mobility on the localization radius and temperature follows the law
\begin{equation} \label{muH-ES}
\mu_{hall} \propto  a^2 \exp\left[ - \alpha_{ES} \left(\frac{T_{ES}}{T}\right)^{1/2}\right],
\end{equation}
where $T_{ES}$ is the temperature from the Efros-Shklovskii law. It is proportional to $a^{-1}$. The numeric value for $\alpha_{ES}$ is $\alpha_{ES} = 0.3$.

\section{Comparison with experiment}
\label{sec_exp}

Our theoretical investigation indicates that Hall effect can be detected more easily in systems with large localization radius. The range of critical exponent values $\xi_c$ accessible in the experiment is limited by the exponential growth of system resistance at large $\xi_c$. For reasonable values of the critical exponent,  the temperature dependence of the Hall mobility converges to the universal VRH curve only for relatively large $n^{1/2}a$. For small $n^{1/2}a$, it converges at very large $\xi_c$. Finally, at the same values of the critical exponent, the system with large $a$ should display a larger Hall effect $\propto a^2$ due to the scaling arguments.

 \begin{figure}[h]
	\centering
	\includegraphics[width=0.35\textwidth]{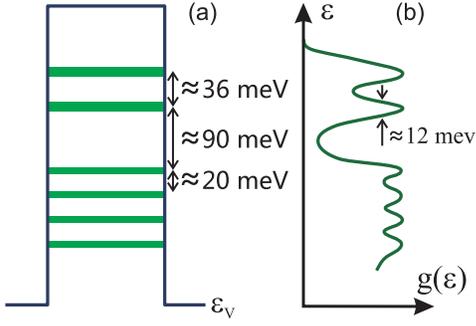}
 	\caption[]{The energy levels of a single quantum dot (a) and the density of states $g(\varepsilon)$ (b). The maximas of $g(\varepsilon)$ correspond to the quantum dot levels. The levels are broadened due to the random potential. }
 	\label{fig:dot}
\end{figure}

As we discussed it in Introduction,  several experimental measurements of the Hall effect were obtained in 3D systems in the vicinity of the metal-insulation transition where the localization radius diverges. Deep in the strong localization regime, where the  localization radius is small, the Hall effect was usually  not visible.
In this section we compare the theory with our recent experimental measurements of the Hall effect in the $p$-doped two-dimensional arrays of tunnel-coupled Ge/Si quantum dots (QDs). These arrays display the VRH conductivity with the localization radius much larger than its typical value in doped semiconductors. The main part of the presented experimental results were preliminary published in Ref.~\cite{Ste17}.

 \begin{figure}[h]
	\centering
	\includegraphics[width=0.35\textwidth]{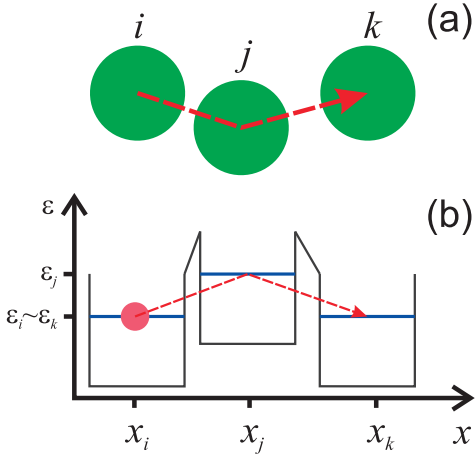}
 	\caption[]{Co-tunneling between distant dots $i$ and $k$. (a) the real space positions of  dots $i$,$j$ and $k$; (b) the  energy diagram where $x$ is the generalized coordinate. }
 	\label{fig:cotun}
\end{figure}

The QD arrays were grown with the low temperature (about of 300$^\circ$C) molecular beam epitaxy. The small size of a quantum dot (lateral size 15-20 nm  and 1.5-2 nm high) leads to a large energy separation of  quantum levels $\gtrsim 10 meV$ and to the nonmonotonic density of states (Fig.~\ref{fig:dot}).
The dominant mechanism of the transport in the discussed arrays is the variable-range hopping between quantum dots. It was shown with the temperature dependencies of conductance to follow the Efros-Shklovskii law $\sigma(T)=\gamma T^m \exp[-(T_{ES}/T)^{1/2}]$. Value $m$ was close to zero~\cite{Yak03}.

The mechanism of the variable range hopping transport in quantum dot arrays is slightly different from the VRH mechanism  in doped semiconductors. The tunneling path to a distant quantum dot inevitably crosses other (intermediate) dots.  The hop to a distant dot includes the co-tunneling process involving the states in the intermediate dots. This process is schematically shown in Fig.~\ref{fig:cotun}. The hole from  site $i$ cannot hop to  site $j$ because of the large energy difference $|\varepsilon_i-\varepsilon_j| \gg T$. Instead, it hops on the distant site $k$ with $\varepsilon_k \sim \varepsilon_i$. The process involves the state on  quantum dot $j$ as the intermediate virtual state. The tunneling amplitude for hop $i\rightarrow k$ can be estimated as $\widetilde{I}_{ik} = I_{ij} I_{jk}/(\varepsilon_i-\varepsilon_j)$. Here $I_{ij}$ is the overlap integral between the states on quantum dots $i$ and $j$. The co-tunneling can involve any number of intermediate dots.

 The transport in QD arrays due to the co-tunneling processes  can be described with the conventional variable range hopping theory \cite{Shk-QD}. However, the localization radius $a$ in the theory is not the localization radius of a single QD, but it is strongly modified by the virtual states on the intermediate dots. This modified localization radius can easily be comparable to or larger than the size of a quantum dot.  It depends on the density of states at the Fermi level~\cite{Nenas,Ste17}. It allows us to control $a$ by changing the QD filling factor $\nu$ and the Fermi level position  with respect to the density of states.

 \begin{figure}[h]
 	\centering
 	\includegraphics[width=0.45\textwidth]{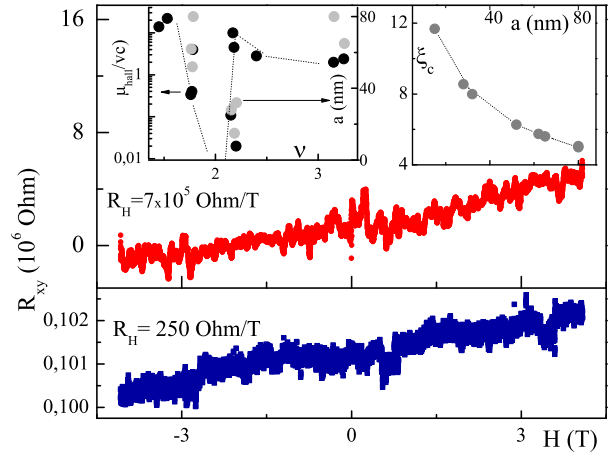}
 	\caption[]{$R_{xy}(H)$ dependence  for two high-resistance samples. Left inset -- the dependence of mobility (black symbols) and localization radius (grey symbols) on filling factor.  Right inset -- $\xi_c(a)$ dependence.}
 	\label{fig:Rxy}
\end{figure}
 	
We measured the Hall effect in the structures with different dot filling factors, which were varied by changing the boron concentration in the  $\delta$-boron-doped silicon layer which is 5 nm below the quantum dot layer.
 The conductance values for the samples under study were shown to be in the range $\sim$10$^{-5}$ -- 10$^{-11}$ Ohm/$\Box$ at 4.2 K. From the comparison of the measured $\sigma(T)$ dependence with the Efros-Shklovskii law we determined the critical exponent $\xi_c = (T_{ES}/T)^{1/2}$ and localization radius  $a=Ce^2/\epsilon k_B T_{ES}$. Here $\epsilon$ is  the permittivity, $k_B$ is the Boltzmann constant, and $C$ is  a numerical coefficient that, according to Ref. \cite{Zig}, is equal to 6.2. It was shown that the localization radius changes from $\sim$25 to $\sim$80 nm depending on the filling factor and,  correspondingly, the Fermi level position.

In Fig.~\ref{fig:Rxy} we show $R_{xy}$ for two high-resistance samples.   The Hall coefficients $R_H$  were determined from the slopes of  $R_{xy}(H)$  lines as $R_H=[R_{xy}(H)-R_{xy}(-H)]/2H$. It allows avoiding the symmetrical contribution of magnetoresistance due to the  asymmetry of the contacts and a possible shift of the amplifier's zero. We observed a strong correlation between  $\mu(\nu)$ and $a(\nu)$ nonmonotonic  dependencies that are obviously observed  in the left inset to Fig.~\ref{fig:Rxy}.

\begin{figure}[h]
	\centering
	\includegraphics[width=0.5\textwidth]{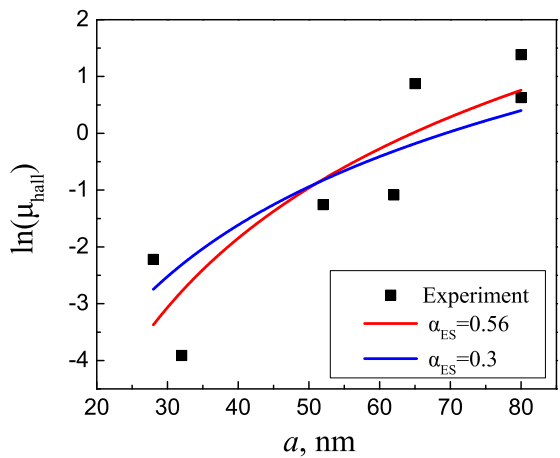}
	\caption[]{The comparison of experimental results with eq.~(\ref{muH-ES}). The red curve corresponds to $\alpha_{ES}$ treated as the fitting parameter while for the blue curve we adopt the value $\alpha_{ES}=0.3$ obtained from the numerical simulation.  }
	\label{fig:compar-exp}
\end{figure}

In Fig.~\ref{fig:compar-exp} we compare the experimental results to eq. (\ref{muH-ES}). The blue curve corresponds to the parameter $\alpha_{ES}=0.3$  obtained from the numerical simulation. The red curve corresponds to $\alpha_{ES}$ treated as the fitting parameter. Expression (\ref{muH-ES}) agrees to the experimental data. The best agreement is achieved for $\alpha_{ES}=0.56$. However, the precision of experimental measurements is insufficient to reliably prove this value. Let us note that there are physical reasons for the measured Hall conductivity to be different from the predictions of the theory. The complex nature of long-range hopping in QD arrays can modify the dependence $\mu_{hall}(a)$ obtained from the point-like site model. However, further experimental investigations are needed to understand if it is the case.

\section{Discussion}
\label{sec_discus}

The conventional approach to the hopping transport in strongly disordered systems is the percolation theory that is based on the assumption that all the exponentially small terms can be neglected at a sufficiently strong disorder. This assumption leads to the result that the Hall effect is controlled by the rare optimal triads of sites. In the NNH regime, they are the triads that form  equilateral triangles with side $r_{perc}$ corresponding to the percolation resistance. The triangles should be positioned in the intersections of three branches of the percolation cluster. In the VRH regime, the restrictions to the optimal triangles are even more solid. They, still, are the triangles consisting of critical resistors. However, now they should include only the critical resistors of the specific kind: the ones with a small length and large energies.

Our numerical results show that the model of optimal triads works well in the NNH regime. The dependence (\ref{muHnn}) obtained from the optimal triangle model agrees to the numerical results. Moreover, the dominant area of the triangle $\langle S_{ikj} \rangle_{hall}$ calculated at a high temperature is in a quantitative agreement with the prediction of the NNH percolation theory $(\sqrt{3}/4)r_{perc}^2$. However, the optimal triangle model fails to describe the VRH case. The calculated temperature dependence of $\mu_{hall}$ follows the exponential laws (\ref{muH-mott}) and (\ref{muH-ES}) instead of the power law predicted by the percolation theory. The clearest evidence of the failure of optimal triangle model is the temperature dependence of $\langle S_{ikj} \rangle_{hall}$. The area of the optimal triangle is smaller in the VRH regime than in NNH. However, $\langle S_{ikj} \rangle_{hall}$ grows with the decreasing temperature, indicating that the Hall effect is controlled by the triangles that are larger even than the optimal triangle of the NNH regime. To our opinion, it indicates that  the Hall effect in the VRH regime is not dominated by rare optimal triangles, but by more numerous ``typical'' triangles. The area of the typical triangle increases with the decreasing temperature because more distant hops become important in the VRH regime.

Why does the optimal triangle model work well in one regime and fail in the other? Here we argue that the reason is the extremely low probability $p_{\triangle}$ to find the optimal triangle of the VRH regime. First, the probability for three branches of the percolation cluster to be connected with a triangle of critical resistors is already small. Each branch of the percolation cluster contains at least one critical resistor. However, it can be positioned in any place of the branch, not necessary at its edge. The branch of the percolation cluster contains $\propto \xi_c$ resistors. Therefore, we estimate the probability to find a critical resistor at the edge of the branch as $1/\xi_c$. The triangle of critical resistors at the intersection of three branches can appear when all the three branches contain critical resistors on their edges, therefore, its probability can be estimated as $1/\xi_c^3$. This estimate can be compared to the numerical result for the pre-exponential part of the dependence (\ref{muHnn}), $(a/r_{perc})^{\gamma_{nn}}\propto \xi_c^{-\gamma_{nn}}$. Note that the critical exponent in the NNH regime is equal to $2r_{perc}/a$. The numerical value $\gamma_{nn}=3.3$ is close to our simplified estimate $3$.

In the VRH regime, the optimization of the exponent (\ref{xi-ikj}) leads to the following restriction to the energies of the sites composing the ``optimal triad''. Two energies should lie in the interval $(\varepsilon_F + (\xi_c-1)T,\varepsilon_F + \xi_c T)$, while the third one can have any energy larger than $\varepsilon_F$ (there is the second possible option, when the three sites have the energies lower than $\varepsilon_F$ and the energies of two sites are in the interval $(\varepsilon_F - \xi_c T,\varepsilon_F - (\xi_c-1) T)$. This option leads to the similar results). If we presume that the distribution of site energies  in the critical resistor is flat, it will add an additional factor $\xi_c^{-2}$ to $p_{\triangle}$ leading to the estimate $p_{\triangle} \propto \xi_c^{-5}$. We argue, however, that this expression overestimates $p_{\triangle}$. There are two reasons for it. The first reason is discussed in Ref. \cite{FP81} and is related to the small probability of finding a site with a maximum possible energy in the percolation cluster.

\begin{figure}[h]
	\centering
	\includegraphics[width=0.3\textwidth]{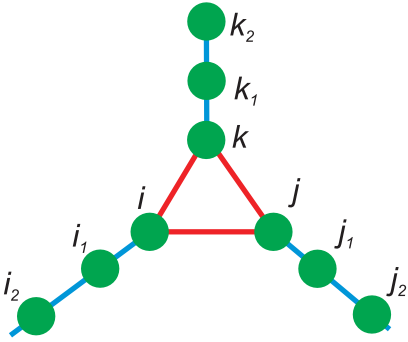}
	\caption[]{The ``optimal triad'' with its connection to the nearest sites of the percolation cluster. The critical resistors
composing the triad are shown with red. Other resistors of the percolation cluster are shown with blue. }
	\label{fig:Ptri}
\end{figure}

Here we want to discuss the second reason that has, to the best of our knowledge, been never discussed. It is related to the fact that the ``optimal triangle'' should be connected to the rest of the percolation cluster, and it is not easy for the considered energies of its sites. In Fig.~\ref{fig:Ptri} we show the optimal triangle with the nearest sites of  three branches of the percolation cluster. For the sake of the qualitative estimate we consider the branches to be straight lines with angle $2\pi/3$ between them. We assume that the two sites of the triangle with energies $\sim \xi_c T$ are $i$ and $j$. The resistors $i-i_1$ and $j-j_1$ are part of the percolation cluster. Therefore, $\varepsilon_{i,i1}/T + 2r_{i,i1}/a \lesssim \xi_c$  and $\varepsilon_{j,j1}/T + 2r_{j,j1}/a \lesssim \xi_c$. The energies in these relations are large $\varepsilon_{i,i1} \ge \varepsilon_i - \varepsilon_F \sim \xi_c T$, $\varepsilon_{j,j1} \ge \varepsilon_j - \varepsilon_F \sim \xi_c T$. It means that  distances $r_{i,i1}$ and $r_{j,j1}$ should be small, of the order of $a/2$. However, resistors $i_1-j_1$, $i_1-k$ and $j_1-k$ should not shunt  resistors $i-j$, $i-k$ and $j-k$. It imposes serious restrictions on energies $\varepsilon_{i1}$ and $\varepsilon_{j1}$.
\begin{equation}\label{restr-i1j1k}
\begin{array}{l}
(\sqrt{3} + 1) + \max(\varepsilon_{i1}-\varepsilon_F,\varepsilon_{j1}-\varepsilon_F)/T \gtrsim \xi_c, \\
\sqrt{3}  + \max(\varepsilon_{i1}-\varepsilon_F,\varepsilon_{k}-\varepsilon_F)/T \gtrsim \xi_c, \\
\sqrt{3} + \max(\varepsilon_{j1}-\varepsilon_F,\varepsilon_{k}-\varepsilon_F)/T \gtrsim \xi_c. \\
\end{array}
\end{equation}
Here we assumed that the sides of triangle $ijk$ are equal to $a/2$. These restrictions mean that, in addition to energies $\varepsilon_i$, $\varepsilon_j$, two other energies from $\varepsilon_{i1}, \varepsilon_{j1},\varepsilon_k$ should also be close to $\varepsilon_F + T\xi_c$. However the range of possible energies $\sim(\varepsilon_F + \xi_cT - (\sqrt{3}+1)T,\varepsilon_F + \xi_c T)$ is larger than the one for $\varepsilon_i,\varepsilon_j$. If the critical exponent $\xi_c$ is large compared to $\sqrt{3}+1$, another similar arguments lead to the restrictions for the energies of other sites $\varepsilon_{k1}$, $\varepsilon_{i2}$, $\varepsilon_{j2}$ etc. It means that the ``optimal triangle'' of the VRH percolation theory is actually quite a  sophisticated and improbable complex of sites that allow the connection of the triangle to the rest of the percolation cluster. The number of the sites in this complex grows with $\xi_c$ leading to the dependence $p_{\triangle}(\xi_c)$ that is stronger than any power law.

However, even if we consider only the restrictions (\ref{restr-i1j1k}),  the apparent dependence $p_{\triangle}(\xi_c)$ is $p_{\triangle} \propto \xi_c^{-7}$. These small probabilities should be compared to the contribution of the nonoptimal triangles that can be evaluated as $\exp(-\alpha_{Mott}\xi_c)$ from equation (\ref{muH-mott}) with $\alpha_{Mott} = 0.47$. Condition $\exp(-0.47\xi_c) < \xi_c^{-7}$ yields $\xi_c > 60$. These values can hardly be accessible in experiment. Note that the dependence of sample resistance on $\xi_c$ is $R=R_0\exp(\xi_c)$. If $R_0\sim 1\,Ohm$, $\xi_c = 60$ leads to $R>10^{26}\,Ohm$. It means that at measurable system resistances, the optimal triads should not be important for the Hall effect in the VRH regime.

In a wide range of temperatures and localization distances the Hall mobility can be described with the law $\mu_{hall} \propto a^2\exp(-\alpha \xi_c)$. Let us note that the range of possible $\xi_c$ values is limited by the conditions of the VRH regime and the restriction for reasonably large conductivity. It makes the structures with the hopping transport and large localization radius $a$ a good choice to study the Hall effect in the VRH regime. The localization radius in the structures discussed in Sec.~\ref{sec_exp} is $\sim 10$ times larger than in typical doped semiconductors. It means that the Hall effect in these structures should be $\sim 100$ times larger than in ordinary semiconductors at the same $\xi_c$. However, the complex nature of  long-range hops in these structures (that inevitably includes co-tunneling) can modify the physics of Hall effect. The comparison of the present theory with experiments in QD ensembles with the VRH transport shows that eq. (\ref{muH-ES}) describes the experimental data. Nevertheless, the $\alpha_{ES}$ values  obtained from the experiment and from numerical simulations of the point-like site model with the Coulomb gap do not seem to be different.

Finally,  we want to note that the Hall effect is not the only phenomena in the hopping transport that is related to the two-phonon hopping with interference. Recently, it was shown that the current-induced spin polarization, spin galvanic effect and spin Hall effect also appear due to the similar processes \cite{Smi-Gol}. The effects were controlled by the complex interplay of the disorder strength and spin relaxation time relation to the hopping time. However, the theory was made only for the case of position disorder, i.e. the NNH regime. We argue that the theory \cite{Smi-Gol} can be reduced to the theory of ordinary Hall effect in the limit of large spin relaxation time. Therefore, our results on the optimal triangles should be important for the theory of spin generation, at least, in some limiting cases.

In conclusion, we revised the theory of the hopping Hall effect in 2D systems. We compared the predictions of the percolation theory to the numerical simulations based on the solution of modified Kirchhoff equations in different regimes. The percolation theory is in agreement with the numerical results in the neighbor hopping regime. However, in the variable hopping regime,  it fails to describe the results of simulation. We argue that it is related to the extremely small probability of finding the optimal triad of the sites in the VRH regime due to the complex nature of the triad and its connection to the percolation cluster. The numerical results in the VRH regime can be described by an empirical law that agrees to our recent experimental data.

The authors are grateful to D.S. Smirnov, Y.M. Galperin, V.I. Kozub and A.V. Nenashev for many fruitful discussions.
N.P.S. acknowledge the support from RFBR (Grant ¹ 16-02-00553).

\appendix

\section{Derivation of the hopping rates}
\label{App1}

First, we derive the rates of the ordinary two-site hops with one phonon. It is instructive to explain our method for this simple case. The ordinary hops are described by the reduction of  second-order density matrix $\langle \widehat{\rho}^{(2)} \rangle_i$. We describe here the hopping between sites $i$ and $j$. It appears in the second order perturbation over $T_{ij}$. It means that the states of all the sites other than $i$ and $j$ are not modified during the hop. Therefore we can start from the density matrix $\widehat{\rho}^{(0)}$ already reduced over all the sites $k \ne i,j$. We express this density matrix as
\begin{multline}\label{A1}
\widehat{\rho}^{(0)} = f_i f_j a_i^+ a_j^+ | \emptyset\rangle \langle \emptyset | a_j a_i +
f_i (1-f_j)a_i^+ | \emptyset\rangle \langle \emptyset | a_i + \\
f_j(1-f_i) a_j^+ | \emptyset\rangle \langle \emptyset | a_j +
(1-f_i)(1-f_j)| \emptyset\rangle \langle \emptyset |.
\end{multline}
Here $| \emptyset\rangle \langle \emptyset |$ describes the ``vacuum'' state when  sites $i$ and $j$ are free and phonons are in the equilibrium.

The hopping is possible only for the second and the third term in (\ref{A1}), i.e. when the two sites have one electron. In this case, one can use one electron notation $| i \rangle = a_i^+ | \emptyset \rangle$. Let us consider the hopping from site $j$ to site $i$. To find its rate we should consider the term $f_j (1-f_i) | j \rangle \langle j | $ and find the contribution of $\langle \rho^{(2)}\rangle_i$ to the state with filled site $i$. It allows one to give the rate of hopping $j\rightarrow i$ as
\begin{multline}\label{A2}
W_{ij} =   - \frac{1}{\cal T} \times \\  {\rm Tr}_{\rm ph}\int_{-\cal T}^{\cal T} dt_1  \int_{-\cal T}^{t_1} dt_2
\langle i|
\left[T_{ij}(t_1),\bigl[T_{ij}(t_2), | j \rangle \langle j | \bigr]\right] |i\rangle.
\end{multline}
Here ${\rm Tr}_{\rm ph}$ is the trace over the final states of the phonon subsystem. ${\cal T}$ is some time interval that is large compared to $\varepsilon_i-\varepsilon_j$. The expression under the integral depends only on the difference $t_1 - t_2$. Therefore, the average time $(t_1 + t_2)/2$ can be integrated out and it is canceled with the time interval ${\cal T}$.
The operator $T_{ij} = t_{ij}a_i^+ a_j + t_{ji}a_j^+a_i$ corresponds to the transition of the electron between sites $i$ and $j$ with the simultaneous emission or absorbtion of some number of phonons. Let us denote the full energy of the system when the electron is on site $j$ as $E_j$. Different $E_j$ values  are possible due to the fluctuations of phonon numbers.
\begin{multline}\label{A3}
W_{ij} =   t_{ij} t_{ji} {\rm Tr}_{\rm ph} \left( \Phi_{ij}  \rho_{ph}^{(0)} \Phi_{ji}   \int e^{i(E_i-E_j)t'} dt' \right) .
\end{multline}

\begin{widetext}
The integral over time $t' = t_1-t_2$  yields the delta-function $\delta(E_i - E_j)$ indicating the energy conservation during the hop.
The term ${\rm Tr}_{\rm ph}\Phi_{ij} (t_1) \rho_{ph}^{(0)} \Phi_{ji} (t_2)$, where $\rho_{ph}^{(0)}$ is the equilibrium phonon density matrix contains the electron-phonon interaction and the probability of finding the phonons required for the hop.
If we consider the electron-phonon interaction to be small, the one-phonon processes will dominate the hop, and we find
\begin{equation}\label{A4}
{\rm Tr}_{\rm ph} \left[ \delta(E_i-E_j) \Phi_{ij}  \rho_{ph}^{(0)} \Phi_{ji}\right] = 2 |\gamma(q)|^2 g^{(ph)}(q)\left(\theta(\varepsilon_j - \varepsilon_i) + \frac{1}{e^{|\varepsilon_j - \varepsilon_i|/T} -1} \right).
\end{equation}
Here $g^{(ph)}(q)$  is the density of states for the phonons with $q = |\varepsilon_i - \varepsilon_j|/c$, where $c$ is the sound velocity. $|\gamma(q)|^2$ describes the strength of the  electron-phonon coupling at the given absolute value of wavevector $q$. In the case of large energies, $\varepsilon_i,\varepsilon_j \gg T$, eq. (\ref{A4}) can be estimated as
\begin{equation}\label{A5}
{\rm Tr}_{\rm ph} \left[\delta(E_i-E_j) \Phi_{ij}  \rho_{ph}^{(0)} \Phi_{ji}\right] \approx \frac{1}{2\pi t_0^2 \tau_1}{\cal N}(\varepsilon_j- \varepsilon_i), \quad
{\cal N}(\varepsilon_j- \varepsilon_i) = \left\{
\begin{array}{lll}
1, & \, & \varepsilon_j > \varepsilon_i, \\
e^{(\varepsilon_j-\varepsilon_i)/T}& \, & \varepsilon_j < \varepsilon_i.
\end{array}
\right.
\end{equation}
Here $\tau_1$ is the time constant associated with normal hops. We neglect the possible power law dependence of $\tau_1$ on the site energies in comparison with the strong dependence ${\cal N}(\varepsilon_j- \varepsilon_i)$.  The normal hopping rate can be then expressed as
\begin{equation}\label{A5}
W_{ij} = \frac{1}{\tau_1} \exp(-2r_{ij}/a) {\cal N}(\varepsilon_j - \varepsilon_i).
\end{equation}
The quantum mechanics of the hop is illustrated in the diagram shown in Fig.~\ref{fig:1phon}. The two solid lines correspond to the two electron density matrix indexes. The interaction with phonons is shown with the ring. Note that, after the polaron transformation, any number of dashed phonon lines can enter the vertex. The energy should be conserved in the vertex. The phonon line corresponds to the factor ${\cal N}(\varepsilon_j - \varepsilon_i)$.

\begin{figure}[htbp]
    \centering
        \includegraphics[width=0.2\textwidth]{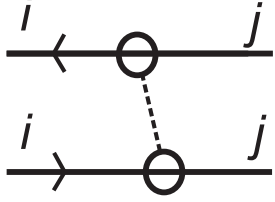}
        \caption{One-phonon processes that leads to ordinary hops. }
    \label{fig:1phon}
\end{figure}

The Hall effect in the hopping transport cannot be described with the ordinary hops (\ref{A5}). Its description should include  the hops $j \rightarrow i$ that occur in the presence of the third site $k$. Let us first consider the situation when  site $k$ is free. The corresponding hopping rate $W_{ikj}^{(0)}$ can be expressed as
\begin{equation} \label{A2-1}
W_{ikj}^{(0)} = i \frac{1}{{\cal T}} {\rm Tr}_{\rm ph}
\int_{t_1 >t_2>t_3}
\langle i|
\left[T_{ikj}(t_1),\left[T_{ikj}(t_2),\bigl[T_{ikj}(t_3), | j \rangle \langle j | \bigr]\right] \right]
 |i\rangle.
\end{equation}
Here $T_{ikj} = T_{ik} + T_{ij} + T_{jk}$. Although the commutators in (\ref{A2-1}) contain quite a number of terms, only
some of them lead to the final state of the electron on site $i$ and yield the non-zero contribution to $W_{ikj}^{(0)}$. We separate the relevant terms
\begin{equation} \label{A2-2}
W_{ikj}^{(0)} = i \frac{1}{{\cal T}} {\rm Tr}_{\rm ph}
\int_{t_1, t_2>t_3}
\langle i|
\Bigl( T_{ij}(t_1)| j \rangle \langle j | T_{jk}(t_3) T_{ki}(t_2) - T_{ik}(t_2) T_{kj}(t_3)| j \rangle \langle j | T_{ji}(t_1) \Bigr)
 |i\rangle.
\end{equation}
Note the difference between the time integration between expressions (\ref{A2-1}) and (\ref{A2-2}). In (\ref{A2-2}) the relation between $t_1$ and other times is arbitrary. The time integration in the first term of expression (\ref{A2-2}) yields
\begin{equation} \label{A2-2-2}
2\pi \delta(E_i - E_j) \left( \frac{-i}{E_j - E_k} + \pi\delta(E_j-E_k)\right).
\end{equation}
The result of the time integration in the second term is the complex conjugate of (\ref{A2-2-2}). Let us note that the contribution of the terms including $\delta(E_i - E_j)/(E_j-E_k)$ to the hopping rate is proportional to $t_{ij} t_{jk} t_{ki} + t_{ik}t_{kj} t_{ji} = 2 {\rm Re}(t_{ij} t_{jk} t_{ki})$. This contribution has only the quadratic dependence on the applied magnetic field. Although this contribution determine the interference magnetoresistance, in the theory of the Hall effect that is linear on the applied field, we neglect this term. The part of $W_{ikj}^{(0)}$ that is responsible for the Hall effect can be expressed as
\begin{equation} \label{A2-3}
W_{ikj}^{(0)} = -4\pi^2 {\rm Im}(t_{ij}t_{jk}t_{ki}) {\rm Tr}_{\rm ph} \left[ \delta(E_i-E_j)\delta(E_j -E_k) \Phi_{ij}
\rho_{ph}^{(0)} \Phi_{jk} \Phi_{ki} \right].
\end{equation}
The two delta-functions in (\ref{A2-3}) mean that the process should include at least two phonons. There are three two-phonon processes that allow the ``energy conservation laws'' $\delta(E_i-E_j)\delta(E_j -E_k)$. These processes are shown schematically in Fig.~\ref{fig:2phon}. These three processes lead to different  phonons participating in the hop and, correspondingly, to different terms ${\cal N}$.

\begin{figure}[htbp]
    \centering
        \includegraphics[width=0.4\textwidth]{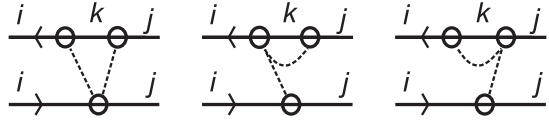}
        \caption{Two-phonon processes that lead to the Hall effect. }
    \label{fig:2phon}
\end{figure}

The sum over possible processes leads to the following expression for $W_{ikj}^{(0)}$
\begin{equation} \label{A2-4}
W_{ikj}^{(0)} = \frac{1}{4} |t_{ij} t_{jk} t_{ki}| \frac{{\bf B} {\bf S}_{ikj}}{2\Phi_0}
\left( \frac{W_{kj} W_{ik}}{|t_{kj}^2 t_{ik}^2|}
+\frac{W_{ij} W_{kj}}{|t_{ij}^2 t_{kj}^2|}
+ \frac{W_{ij} W_{ki}}{|t_{ij}^2 t_{ik}^2|}
 \right).
\end{equation}

Now let us discuss the situation when  site $k$  is filled before the hop. In this case, we start from  state $|kj\rangle = a_k^+ a_j^+ | \emptyset \rangle$. Let us note that operators $a_k^+$ and $a_j^+$ anticommute and, in our notations, $|kj\rangle = - |jk\rangle$. The hopping that includes site $k$ corresponds to the electron transition   from site $k$ to site $j$ and the following electron transition of  from site $j$ to site $k$. The hopping rate $W_{ikj}^{(1)}$ can be expressed as
\begin{equation} \label{A3-1}
W_{ikj}^{(1)} = i \frac{1}{{\cal T}} {\rm Tr}_{\rm ph}
\int_{t_1, t_2>t_3}
\langle ik|
\Bigl( T_{ij}(t_1)| kj \rangle \langle jk | T_{ki}(t_3) T_{jk}(t_2) -  T_{kj}(t_2) T_{ik}(t_3)| kj \rangle \langle jk | T_{ji}(t_1) \Bigr)
 |ki\rangle.
\end{equation}
Note the difference in the order of the transitions between expressions (\ref{A3-1}) and (\ref{A2-2}). In (\ref{A3-1}) transition $j\rightarrow k$ occurs after $k \rightarrow i$. It leads to somewhat different averaged phonon numbers and to the expression
\begin{equation} \label{A2-4}
W_{ikj}^{(1)} = -\frac{1}{4} |t_{ij} t_{jk} t_{ki}| \frac{{\bf B} {\bf S}_{ikj}}{2\Phi_0}
\left( \frac{W_{kj} W_{ik}}{|t_{kj}^2 t_{ik}^2|}
+\frac{W_{ij} W_{jk}}{|t_{ij}^2 t_{kj}^2|}
+ \frac{W_{ij} W_{ik}}{|t_{ij}^2 t_{ik}^2|}
 \right).
\end{equation}

\end{widetext}


\begin{thebibliography}{99}

\bibitem{Holstein}
T. Holstein,  Phys. Rev. 124, 1329 (1961)

\bibitem{NSS}
V.L. Nguyen, B.Z. Spivak, and B.I. Shklovskii, Pis'ma Zh. Eksp.
Teor. Fiz. {\bf 41}, 35 (1985); Zh. Eksp. Teor. Fiz. {\bf 89}, 1770
(1985) [Sov. Phys. JETP {\bf 62}, 1021 (1985)]

\bibitem{SS} Shklovskii B.I., Spivak B.Z. In: Hopping
transport in solids, ed. by M.Pollak and B.Shklovskii, Elsevier,
1991, p. 271

\bibitem{our-int}
A.V. Shumilin, V.I. Kozub, Phys. Rev. B {\bf 85}, 115203 (2012)

\bibitem{Mil-Abr}
A. Miller, E. Abrahams, Phys. Rev. {\bf 120} 745 (1960)

\bibitem{ES}
B. I. Shklovskii and A.L. Efros, "Electronic Properties of Doped
Semiconductors" (Springer, Berlin, 1984).


\bibitem{bryksin}
H. Bottger, V.V. Bryksin, phys. stat. sol. B {\bf 81}, 433 (1977)


\bibitem{FP78}
L. Friedman, M. Pollak, Phyl. Mag. B, {\bf 38} 173 (1978)

\bibitem{Galp}
Y.M. Galperin, E.P. German, V.G. Karpov, Sov. Phys. JETP, {\bf 72}, 193 (1991)


\bibitem{Mott-orig}
N.F. Mott, Journal of Non-Crystalline Solids, {\bf 1} 1 (1968)

\bibitem{Ambeo}
V. Ambegaokar, B.I. Halperin, J.S. Langer, Phys. Rev. B {\bf 4}, 2612 (1971)



\bibitem{gru81}
``The hopping Hall mobility - percolation approach'',
M. Gruenewald, H. Mueller, P. Thomas, D. Wuertz, Solid State Communications {\bf 38} lOll (1981)


\bibitem{FP81}
L. Friedman, M. Pollak, Phil. Mag. B, {\bf 44} 487 (1981)

\bibitem{Fri82}
L. Friedman, Phys. Rev. B {\bf 25} 3512 (1982)



\bibitem{Burkov}
A. A. Burkov, L. Balents, Phys. Rev. Lett. {\bf 91}  057202 (2003)

\bibitem{Ano2}
Xiong-Jun Liu, Xin Liu, J. Sinova, Phys. Rev. B {\bf 84} 165304 (2011)


\bibitem{num1}
P.N. Butcher, J.A. Mcinnes, Phil. Mag. B, {\bf 44:5} 595 (1981)


\bibitem{num2}
P.N. Butcher, J.A. Mcinnes  S. Summerfield, Phil. Mag. B, {\bf 48:6} 551 (1983)



\bibitem{Bryksin-Book}
H. Bottger, V.V. Bryksin, Hopping conduction in solids, Akademie-Verlag Berlin 1985

\bibitem{Koon} D.W. Koon, T.G. Castner, Solid State Communications, {\bf64}, 11 (1987)

\bibitem{Essalen} L. Essaleh, S.M. Wasim, J. Galibert, Materials Letters {\bf 60} 1947 (2006)

\bibitem{Zhang} Y. Zhang, P. Dai, I. Kam, M.P. Sarachik, Phys Rev B {\bf 49} 5032 (1994)

\bibitem{Avdonin} A.Avdonin, P.Skupinski, K.Grasza,  Physica B {\bf 483} 13 (2016)

\bibitem{Kajikawa}Y. Kajikawa,  Phys. Status Solidi C {\bf 14},  1600129 (2017)

\bibitem{Amitay}
M.Amitay, M.Pollak, J. Phys. Soc. Jpn. {\bf 21}, Suppl. 549 (1966)

\bibitem{Klein}R.S. Klein, Phys. Rev.B {\bf 31}, 2014 (1985)

\bibitem{Yak00} A.I. Yakimov, A.V. Dvurechenskii, V.V. Kirienko, Yu.I. Yakovlev, A.I. Nikiforov, C.J. Adkins. Phys. Rev. B {\bf61}, 10868 (2000)

\bibitem{Ste17} N.P. Stepina, A.V. Nenashev, A.V. Dvurechenskii, JETP Letters, {\bf 106}, 308 (2017)

\bibitem{Shk-QD}
J. Zhang,  B.I. Shklovskii, Phys. Rev. B {\bf 70}, 115317 (2004)

\bibitem{Yak03} A.I. Yakimov, A.V. Dvurechenskii,
 A.I. Nikiforov, A.A. Bloshkin, JETP Letters,  {\bf 77} 376 (2003)

 \bibitem{Bar} V. I. Kozub, S. D. Baranovskii, I. S. Shlimak, Solid State Communications {\bf 113}, 587 (2000)

\bibitem {Nenas}A. V. Nenashev, J. O. Oelerich and S. D. Baranovskii. J. Phys.: Condens. Matter {\bf27}, 093201 (2015)

 \bibitem{Zab} A. G. Zabrodskii, K.N. Zinov'eva, Sov. Phys. JETP
 {\bf 59}, 425 (1984)

\bibitem {Zig} D.N. Tsigankov, A.L. Efros, Phys. Rev. Lett. {\bf 88},
 176602 (2002)

 \bibitem{Ste06} N.P. Stepina, A.I. Yakimov, A.V. Nenashev, A.V. Dvurechenski,
 N.A. Sobolev, D.P. Leitao, V.V. Kirienko,
 A.I. Nikiforov, E.S. Koptev, L. Pereira,  M.S. Karmo,
 JETP {\bf 103}, 269 (2006)

\bibitem{Mott1} N.F. Mott, E.A.  Davis, R.A Street, Phil. Mag. {\bf 32}, 961 (1975)

\bibitem{Skal} A.S. Skal, B.I. Shklovskii,   Soviet Phys. Semicond., {\bf 8}, 1029 (1975)

\bibitem{Coulomb-ava}
A.V. Shumilin, Solid State Communications {\bf 183} 51 (2014)

\bibitem{Smi-Gol}
D.S. Smirnov, L.E. Golub, Phys. Rev. Lett. {\bf 118} 116801 (2017)


\end{thebibliography}
\end{document}